# Spontaneous symmetry breaking
# in the finite, lattice quantum sine-Gordon model


S.G. Chung

Department of Physics, Western Michigan University, Kalamazoo, MI 49008-5151



Abstract

The spontaneous breaking of a global discrete translational symmetry in the finite, lattice quantum sine-Gordon model is demonstrated by a density matrix renormalization group. A phase diagram in the coupling constant - inverse system size plane is obtained. Comparison of the phase diagram with a Woomany-Wyld finite-size scaling leads to an identification of the Berezinskii-Kosterlitz-Thouless transition in the quantum sine-Gordon model as the spontaneous symmetry breaking.


The sine-Gordon (SG) model has been basically understood, i.e., the Bethe Ansatz (BA) solution[1] and statistical mechanics,[2] in the attractive regime $\beta^2 < 4\pi$. The repulsive regime $4\pi < \beta^2 < \infty$, however, is still open. The Bethe Ansatz for the massive Thirring (MT) model, which is formally equivalent to the SG model for $\beta^2 < 8\pi$,[3] led to physically undesirable charge (topological) neutral excitations.[4] The quantum inverse scattering method for a lattice SG model with local interaction led to the same difficulty.[5] It is believed that the physical vacuum should be a simple Dirac sea. To avoid the difficulty at the repulsive regime, Luther pointed out the equivalence of the MT model with the spin 1/2 XYZ model through the Jordan-Wigner transformation, and obtained an expression for the soliton mass [cf. Eq (23) in Ref. 6].[6] See also a criticism of Wiegmann[7] on the equivalence between the eight vertex model from which one calculates the energy spectrum of the spin 1/2 XYZ model and the SG model. The instability at $\beta^2 = 8\pi$, however, was not properly resolved. In fact, it was later confirmed through extensive perturbative renormalization group studies of the SG model[7,8] in the context of its near equivalence to the 2D XY model and the associated Berezinskii-Kosterlitz-Thouless (BKT) transition, that the SG model undergoes a BKT transition at $\beta^2 = 8\pi$ in the small mass parameter limit. The precise determination of the BKT transtion point and its universality class was done by Nomura and others in a series of papers. [9] That is, the phase $8\pi < \beta^2$ is massless. A possible dynamical mass generation in the massless Thirring model through a Jordan-Wigner mapping to the spin 1/2 XXZ model was discussed by McCoy and Wu.[10] Notice an important difference between the spin 1/2 XXZ and XYZ models. The former has a massless phase while the latter not. Iwabuchi and Schotte also tried to realize the lattice MT model out of a scaling limit of a six-vertex model.[11] The obtained soliton mass [cf Eq (5.10) in Ref. 11] is different from that of Luther, and the massless phase is neither accounted



for. An effort to cover both the massless phase and the massive phase was due to Dutyshev, and Japaridge et al,[12] a U(1) symmetric isospin massless Thirring model which is equivalent to the Luther-Emery backscattering model.[13] They obtained a BKT-like phase diagram with a dynamical mass generation but *without spontaneous symmetry breaking*, and the same soliton mass in the repulsive regime $4\pi < \beta^2 < 8\pi$ as that of Iwabuchi and Schotte. It is found, however, that the underlying particle spectrum in this model is different from that of the SG model for large momenta. It is precisely this difference which makes the U(1) theory free from the difficulty at $4\pi < \beta^2 < 8\pi$. This fact is also a direct evidence that bosonization[7,13,14] which leads to the SG model is precise only at large space-time separations. In short, a unified theory of the SG model which gives the exact soliton spectrum at $4\pi < \beta^2 < 8\pi$, and the massless phase at $8\pi < \beta^2$ is yet to be constructed. A recent work by Kehrein[15] based on Wegner's flow equation method is a good progress in this direction.

So much for the infinite system. The BKT transition-bearing models, however, suffer a strong finite-size effect arising from the essential singularity, exponential growth of the correlation length near the BKT transition.[16] In particular, the infinite order BKT transition is replaced by a 2-nd order like transition with effective critical coupling constant which depends on the system size logarithmically. Thus in reality, when finite condensed matter systems are analyzed by the SG model or any other BKT transition-bearing models, the physical quantities of interest will critically depend on the system size.

It is also worth mentioning that often in condensed matter physics, there exists a physically meaningful lattice cutoff and the lattice cutoff related ambiguity, particularly divergencies and necessary renormalization procedure do not exist. Thus, our first motivation in this paper is to



precisely analyze a finite, lattice SG Hamiltonian,

$$H_\ell = \sum_{i=1}^{L} \left\{ -\frac{\beta^2}{2} \frac{d^2}{d\phi_i^2} + \frac{1}{2\beta^2} (\phi_i - \phi_{i+1})^2 + \frac{1}{\beta^2} (1 + \cos \phi_i) \right\} \qquad (1)$$

where $\phi_i$ is the field variable at the lattice site i. The field theory SG Hamiltonian[17]

$$H_f = \int \left\{ \frac{1}{2} \pi^2 + \frac{1}{2} \phi_x^2 + \frac{m^2}{\beta^2 \ell^2} (1 + \cos \beta \phi) \right\} dx \qquad (2)$$

where $\ell$ is the lattice cutoff, can be written, after discretization and rescaling $\beta\phi \rightarrow \phi$ and m=1, as

$$H_\ell = H_f \ / \ \ell \qquad (3)$$

To discuss our second motivation, consider the strong coupling limit $\beta \rightarrow \infty$ with the infinite system size L $\rightarrow \infty$. It is clear from (2) that in this limit $H_f$ becomes a massless scalar field theory with nondegenerate ground state. In the weak coupling limit $\beta = 0$ on the other hand, the ground state is infinitely degenerate with each ground state describing a zero-point motion near one of the potential minima $\beta\phi = \pi \pm 2\pi\times$integer. The latter point may be intuitively understood if one regards (1) as describing a system of torsion-coupled quantum pendula under gravity. In this picture, $\beta=0$ means an infinite mass of pendulum. One thus expects a quantum phase transition at some critical coupling constant $\beta_c^2$ separating a gapless nondegenerate ground state and a broken symmetry ground state which is simply a zero-point motion. Is the BKT transition in the SG model the spontaneous breaking of a gloval discrete translational symmetry in the $\phi$ space?



In this paper, using a density matrix renormalization group (DMRG)[18], we demonstrate the spontaneous symmetry breaking (SSB) in the finite, lattice SG model.  We draw a phase diagram in the $\beta^2$ - inverse system size plane, a critical line separating the SSB ground state and unbroken one.  Comparing the rssult with a Roomany-Wyld finite-size scaling[19] leads then to the identification of the BKT transition as the SSB.

To analyze $H_\ell$ by DMRG, we proceed as follows.  First determine the basis states at each lattice site by solving the 1-body problem, the Mathieu equation[20]

$$\left\{ -\frac{\beta^2}{2} \frac{d^2}{d\phi^2} + \frac{1}{\beta^2} (1 + \cos \phi) \right\} \psi (\phi) = \varepsilon \ \psi (\phi) \tag{4}$$

To solve this, we limit the $\phi$ space to be $[-M\pi, M\pi]$ and take M to be an even integer.  Then in the Floquet's solution

$$\psi_{n\nu} (\phi) = e^{i\nu\phi} \ P_{n\nu} (\phi) \tag{5}$$

where n is the band index and $\nu$ is crystal momentum, $\nu$ is determined from the periodic boundary condition $e^{2\pi M\nu i} = 1$.  $P_{n\nu} (\varphi)$ is $2\pi$ periodic and can be expanded with a sufficiently large integer J as

$$P_{n\nu} (\phi) = \sum_{k=-J}^{J} C_{n\nu}^k \ e^{ik\phi} \tag{6}$$

It is convenient to work on the Wannier functions



$$W_n \ (\phi - 2\pi m) = \frac{1}{\sqrt{M}} \ \sum_\nu \ e^{-2\pi m\nu i} \ \psi_{n\nu} (\phi) \qquad\qquad (7)$$

where $m = -\dfrac{M-1}{2} \ , -\dfrac{M-1}{2} + 1 , \ldots , \dfrac{M-1}{2}$ . The Wannier function is localized at each cosine

potential well. Including up to n bands and for fixed number of $\nu$ states, M, the dimension of the

local basis states is q = n x M and all the local variables are expressed by q x q matrices.

To calculate the ground state and the first excited state, and thus an energy gap Gap(L) as a

function of the system size L, we follow the standard DMRG procedure. We use the infinite

algorithm, open boundary condition, and the ground state target. We limit the phase space at each

lattice site to 4 potential wells, i.e. M = 4. We put n = 4 and start with the superblock size N = 40.

The cases n = 5 and N = 45 are checked for the case $\beta^2$ =13 to see the convergence. The superblock

sizes N=50 and 60 are also checked for the cases $\beta^2$ =16-18

Figures 1-3 are the results for $\beta^2$ = 13. Fig. 1 shows the probability distribution of the phase

(position of pendulum in mechanical analog) at the center site in the ground state. The probability

distributions at different sites differ only a few % at the edges. Due to the phase space truncation,

M = 4, the translational symmetry is somewhat broken from the outset, and the symmetry unbroken

state at L = 7 is delocalized over the two potential minima at -$\pi$ and $\pi$. With the increase of the

system size, the distribution becomes asymmetric and eventually localized near the potential well

at -$\pi$. At the same time, the first excited state shows similar localization but at the other potential

minimum at $\pi$. At L = 43, the two states are almost degenerate, the energy difference ~$10^{-5}$, showing

the SSB and the associated ground state degeneracy. Fig. 2 shows the phase averages at the center

site for the lowest 2 states as functions of the system size L. After L = 43, the first excited state



suddenly acquires a mass, indicating that the $+\pi$ ground state localized at the potential well at $+\pi$ is no more accessible from the $-\pi$ ground state, and the excited state thereafter is due to a local deformation of the $-\pi$ ground state which must be a topologically neutral soliton-antisoliton pair creation. The squares in Fig. 3 show the phase average, now different from site to site, vs the lattice site in the first excited state at $L = 67$.

We repeat the calculation varying the coupling constant $\beta^2$. With the decrease of $\beta^2$, the SSB occurs for shorter system sizes, more abruptly, and the soliton-antisoliton pair becomes more deeply bounded as shown in Fig.3. We can now draw a phase diagram in $\beta^2$ - L plane with a critical line separating the broken symmetry ground state and the unbroken one. To clearly see an asymptotic behavior at large system size, we have rather plotted the $\beta^2$ - 1/L phase diagram in Fig. 4. In this figure, a simple extrapolation from the last three points for the critical coupling constants $\beta_c^2 = 16\text{-}18$ gives $\beta_c^2 = 19.0$ at $L \rightarrow \infty$. This value is different from the well-established $\beta^2 = 8\pi$ for the BKT transition in the small mass limit $m \rightarrow 0$ (cf. (2)). However, we have to take into account the fact that we made the limited phase space approximation, and that the BKT-bearing systems suffer a strong finite-size effect.[16] The infinite order BKT transition is replaced by a 2-nd order transition with logarithmically size-dependent critical coupling constant. We thus need to evaluate the critical coupling constant associated with the finite-size modified, and limited phase space modified, BKT transition. For this purpose, the Roomany-Wyld finite size scaling[19] tells us that a phase transition can be identified by measuring the quantity L x Gap (L) vs $\beta^2$. Fig. 5 shows L x Gap (L) vs $\beta^2$. The data crossing at $\beta_c^2 = 18.8$ indicates a continuous transition. Note that the situation is rather like the Ising model, where a spontaneous symmetry breaking separates 2 massive phases.[17] From the RG studies of the continuous model (2),[7,8] it is known that the critical point $\beta_c^2 = 8\pi$ in the small mass



limit separates the massless phase and the massive phase. In our lattice model, due to the truncation of the phase space to [-M$\pi$, M$\pi$], the massless phase becomes massive, and therefore the behavior of L x Gap (L) vs $\beta^2$ should look like that of the Ising model. The good agreement between the two values $\beta_c^2$ = 18.8 and 19.0 indicates that the BKT transition is indeed the SSB

To summarize, we have demonstrated the spontaneous symmetry breaking in the finite, lattice quantum sine-Gordon model by using density matrix renormalization group. A phase diagram in the coupling constant - inverse system size plane is obtained. Combining the phase diagram with the Woomany-Wyld finite-size scaling, we have identified the Berezinskii-Kosterlitz-Thouless transition in the quantum sine-Gordon model as the spontaneous symmetry breaking.

I thank David Kaup for his correspondence. This work was partially supported by NSF under DMR990002N and utilized the SGI/CRAY Origin2000 at the National Center for Supercomputing Applications, University of Illinois at Urbana-Champaign.

**Figure Captions**

Fig. 1    The probability distribution of the phase at the center site in the ground state for the system sizes L = 7, 37, 43 and 61. $\beta^2 = 13$.

Fig. 2    The phase average vs the system size for the lowest 2 states for $\beta^2 = 13$. The phase-average split to $\pm\pi$ and energy degeneracy indicate the spontaneous symmetry breaking.

Fig. 3    Phase averages vs the lattice site in the first excited state for $\beta^2 = 7, 9$, and 13.

Fig. 4    The phase diagram in $\beta^2$ - 1/L plane. An extrapolation in the limit L $\rightarrow \infty$ limit gives $\beta_c^2 = 19.0$. $\beta^2 = 8\pi$ for the BKT transition in the small mass limit is also plotted for comparison.

Fig. 5    L x Gap (L) vs $\beta^2$ for the $(-4\pi, 4\pi)$ phase space. The lines are from cross to square for the system sizes L = 19, 25, 31, 37, 43, 49 and 55. Data crossing occurs at $\beta_c^2 = 18.8$.



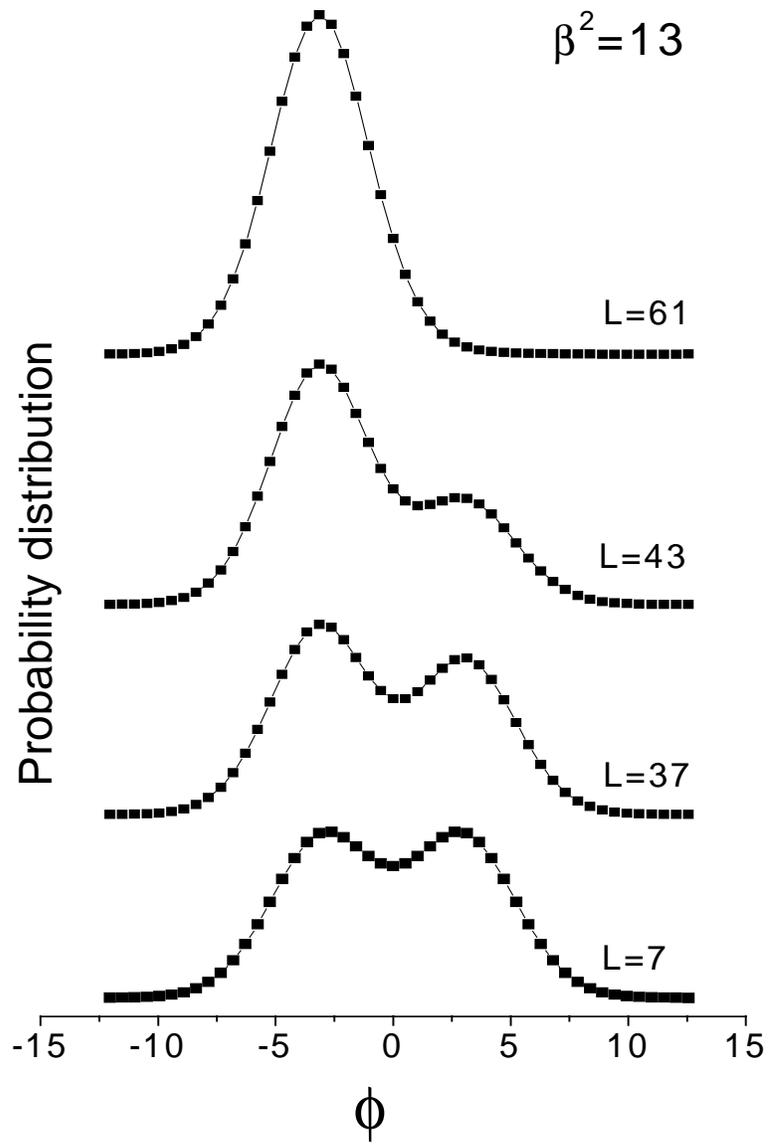

Fig 1



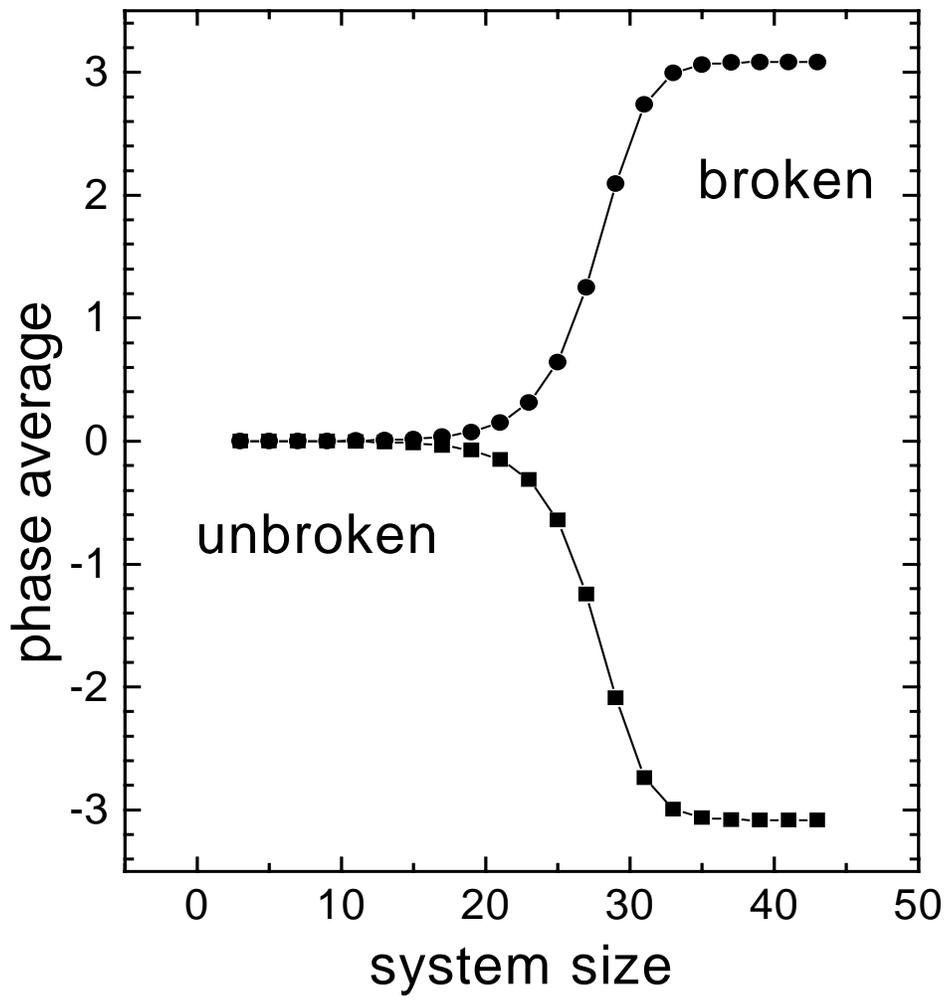

Fig 2



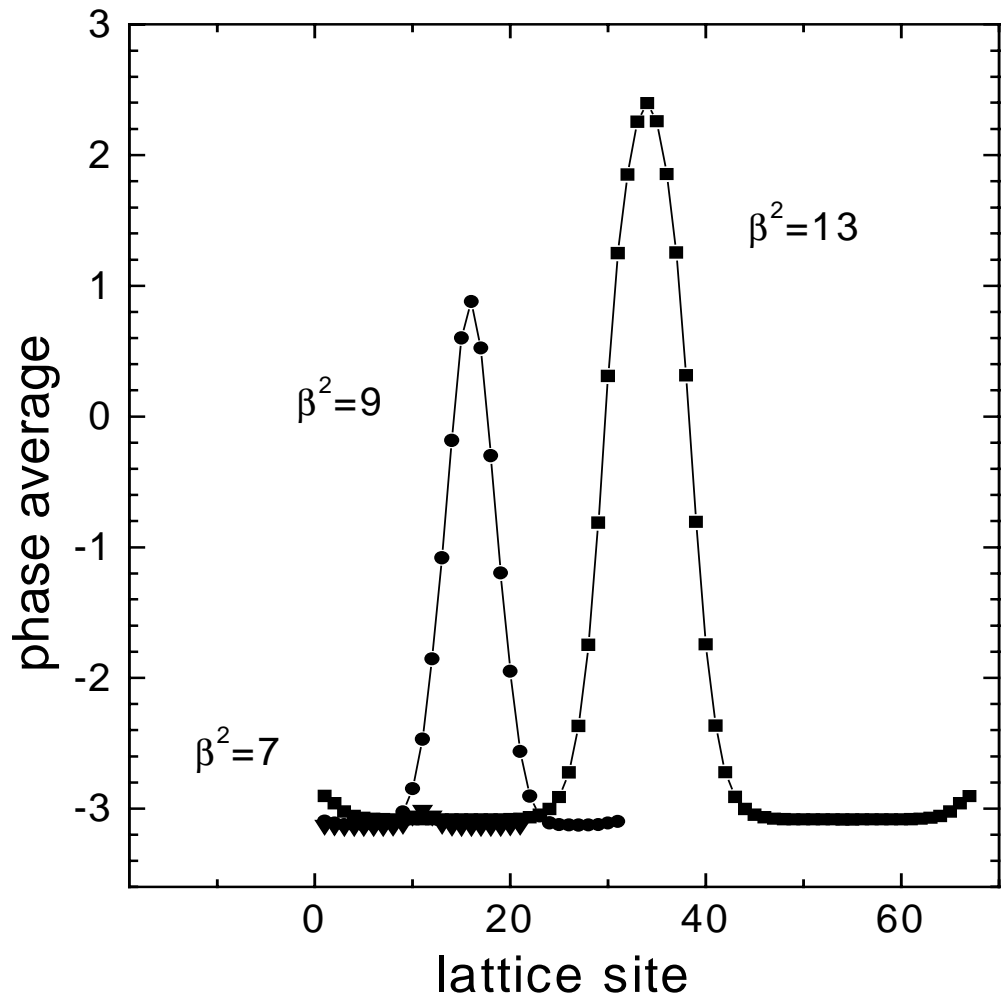

Fig 3



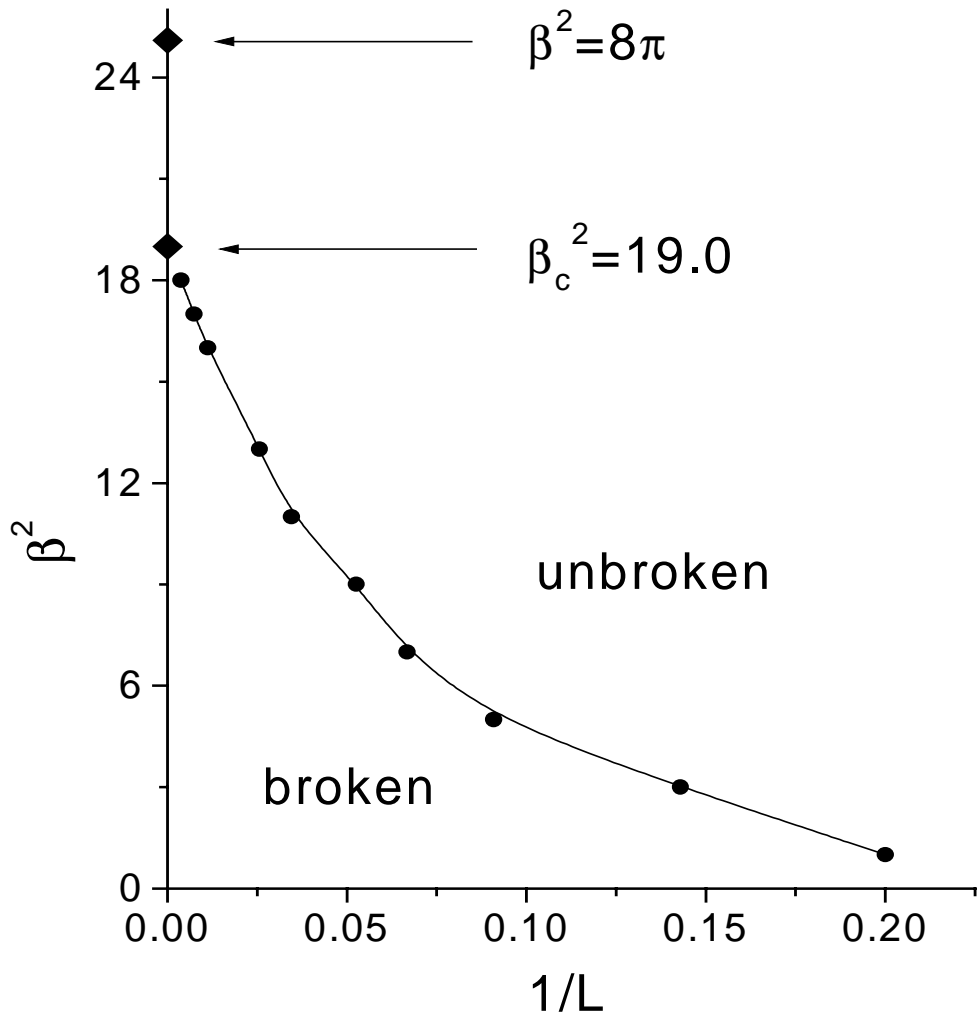

Fig 4



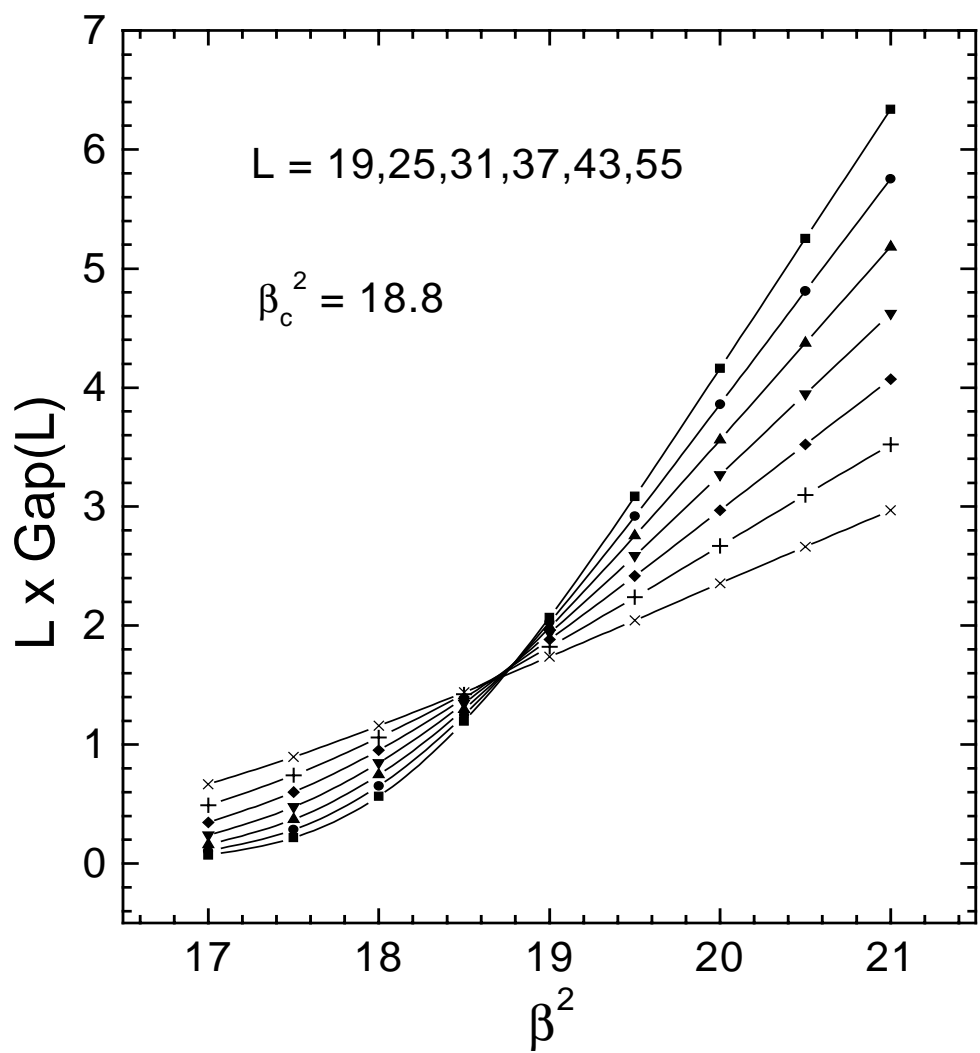

Fig 5